\documentclass[useAMS,usenatbib]{mn2e}
\usepackage{psfig}
\newif\ifAMStwofonts
\newcommand{\lapp}{\mbox{\raisebox{-0.3em}{$\stackrel{\textstyle <}{\sim}$}}}
\newcommand{\gapp}{\mbox{\raisebox{-0.3em}{$\stackrel{\textstyle >}{\sim}$}}}

\title[The double-double radio galaxy 3C293]
      {A radio study of the double-double radio galaxy 3C293}
\author[S. A. Joshi et al.]
       {S. A. Joshi$^1$$\thanks{E-mail: sjoshi@ncra.tifr.res.in (SJ);
                   sumana@aries.res.in (SN);  
                   djs@ncra.tifr.res.in (DJS);
                   ishwar@ncra.tifr.res.in (CHI);
                   chiranjib.konar@gmail.com (CK)}$, 
        S. Nandi$^2$, D.J. Saikia$^1$, C.H. Ishwara-Chandra$^1$ and C. Konar$^3$ \\
$^1$ National Centre for Radio Astrophysics, TIFR, Pune University Campus, Post Bag 3, Pune 411 007, India \\
$^2$ Aryabhatta Research Institute of Observational Sciences, Manora Peak, Nainital 263 129, India \\         
$^3$ Institute of Astronomy and Astrophysics, Academia Sinica, P.O. Box 23-141, Taipei 10617, Taiwan, R.O.C. \\
}
\date{Accepted.                         Received }

\pagerange{\pageref{firstpage}--\pageref{lastpage}}
\pubyear{  }

\begin{document}

\maketitle

\label{firstpage}

\begin{abstract}
We present radio continuum observations at frequencies ranging from 
$\sim$150 to 5000 MHz of the misaligned double-double radio galaxy, DDRG, 
3C293 (J1352+3126) 
using the Giant Metrewave Radio Telescope (GMRT) and the Very Large Array (VLA).
The spectra of the outer lobes and the central source are consistent with being 
straight, indicating spectral ages of $\lapp$17$-$23 Myr for the outer lobes, and
$\lapp$0.1 Myr for the central source. The north-western lobe has a prominent
hotspot suggesting that the interruption of jet activity is $\lapp$0.1 Myr, 
consistent with the age of the inner double. The time scale of interruption of
jet activity appears significantly smaller than observed in most other DDRGs which 
are often associated with giant radio sources. These observations suggest that
there is a wide range of time scales of interruption of jet activity in active 
galaxies.
\end{abstract}


\section{Introduction}
An important aspect in our understanding of active galactic nuclei (AGN) 
is whether their nuclear activity is episodic and if so, the duration of 
their active phases, and the implications of such episodic activity. The
extended radio emission in radio galaxies and quasars contains an imprint
of the history of nuclear jet activity. The structure and spectra 
of their lobes of radio emission provide us with an opportunity 
of studying the time scales of episodic nuclear activity. 
A striking example of episodic nuclear 
activity is when new pairs of radio lobes are seen closer to the nucleus 
before the `old' and more distant radio lobes have faded (e.g. Subrahmanyan, 
Saripalli \& Hunstead 1996; Lara et al. 1999). The ones with two pairs of
radio lobes are usually referred to as `double-double' radio galaxies
(DDRGs; Schoenmakers et al. 2000a); while an example of one with three pairs of 
radio lobes has also been reported (Brocksopp et al. 2007). Presently, close to 
about two dozen good cases of AGN with episodic nuclear activity have been 
identified from either radio and/or X-ray observations (Saikia \& Jamrozy 2009,
for a review). 

Although in most DDRGs, the outer double appears reasonably well aligned 
with the inner one (e.g. Saikia, Konar \& Kulkarni 2006),
3C293 is a striking example of a DDRG where the inner double is misaligned from the
outer one, the misalignment angle being $\sim$35$^\circ$. The radio galaxy 
3C293 is at a redshift of 0.0450 (Fouque et al. 1992) so that 1 arcsec corresponds to 
0.874 kpc in a Universe with H$_0$=71 km s$^{-1}$ Mpc$^{-1}$, $\Omega_{\rm m}$=0.27 and 
$\Omega_{\rm vac}$=0.73.  The optical host galaxy, VV5-33-12, is peculiar
with compact knots and multiple dust lanes (van Breugel et al. 1984; Martel et al. 1999; 
de Koff et al. 2000; Capetti et al. 2000), and appears to be a merger remnant. 
It has a small companion galaxy situated $\sim$37 arcsec ($\sim$30 kpc) towards
the south-west (Heckman et al. 1985; Evans et al. 1999; Beswick et al. 2004). 
Hubble Space Telescope (HST) observations have detected
an optical/infrared (IR) jet within the central kiloparsec emitting
synchrotron emission at these wavlengths (Leahy, Sparks \& Jackson 1999;
Floyd et al. 2006). Emonts et al. (2005) have reported evidence of
jet-induced outflow of warm gas in 3C293.
Broad neutral hydrogen (H{\sc i}) absorption was
observed by  Baan \& Haschick (1981) using the Arecibo telescope,
and has since been studied with a wide range of 
angular resolutions revealing the complex gas distribution seen
in absorption towards the different radio components of the central
source (Shostak et al. 1983; Haschick \& Baan
1985; Beswick et al. 2002, 2004; Morganti et al. 2003). 
Evans et al. (2005) have found CO(1-0) gas in both emission and
absorption within the central few kiloparsecs. The CO emission appears
to be largely distributed in an asymmetric disc rotating about the AGN.
  
%%%%%
\begin{figure*}
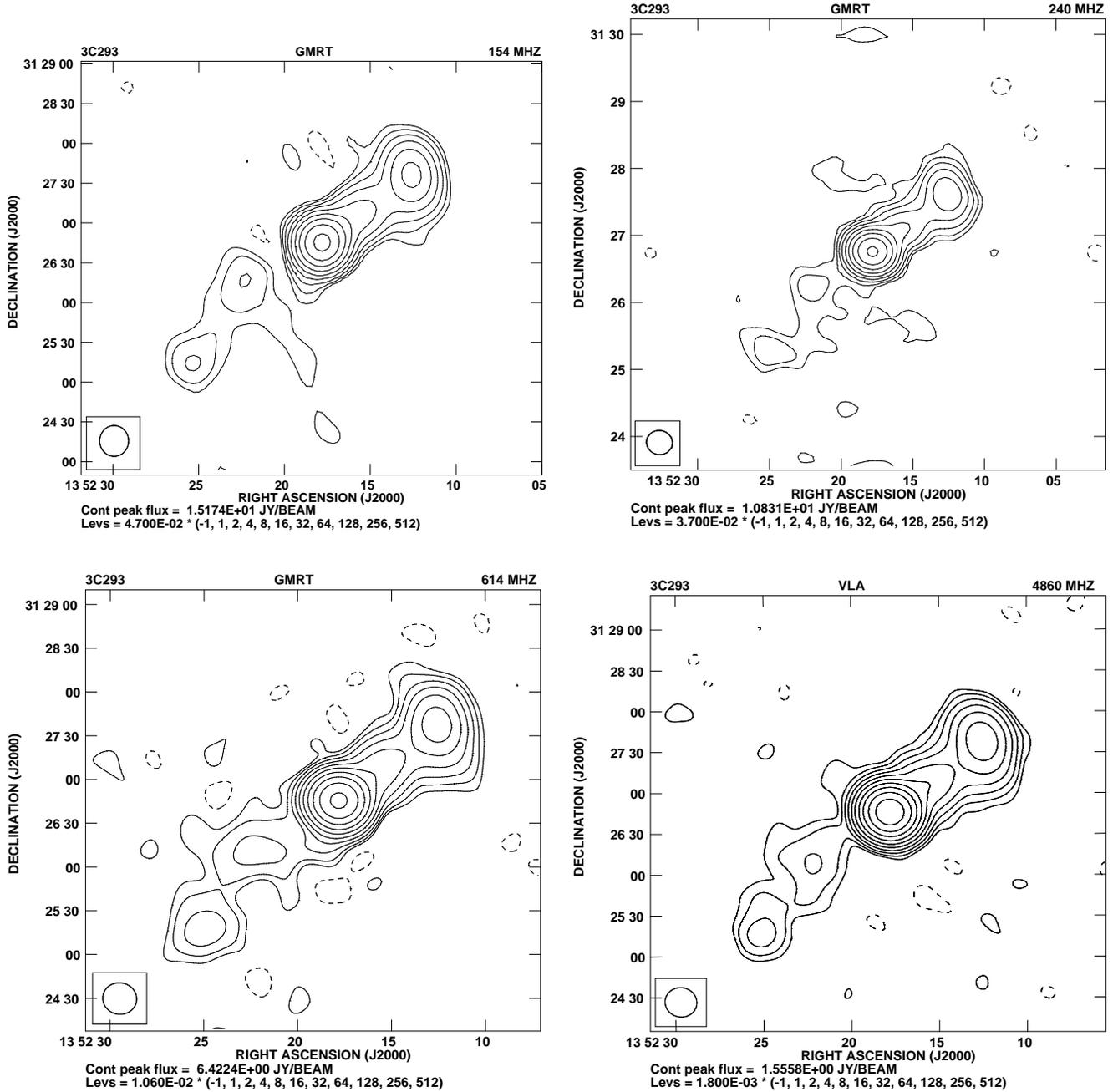

\vbox{
\hbox{
\psfig{file=3C293_150MHZ_PBCOR1.PS,width=3.5in,angle=270}
\psfig{file=3C293_239MHZ.PS,width=3.5in,angle=0}
     }
\hbox{
\psfig{file=3C293_614_NEW.PS,width=3.5in,angle=0}
\psfig{file=3C293_4860MHZ.PS,width=3.5in,angle=-90}
     }
     }
\caption[]{The GMRT images of 3C293 at 154, 240 and 614~MHz, 
and the VLA image at 4860 MHz. 
All these images have been made with an angular resolution
of 23.2$\times$21.6 arcsec$^2$ along a PA of 79$^\circ$, which 
is shown as an ellipse in the bottom left-hand corner.}
\end{figure*}
%%%%%%

The large- and small-scale structure of 3C293 have been imaged by a number 
of authors. The outer double-lobed structure has a projected linear size
of $\sim$190 kpc, and resembles an FRII radio galaxy. However, the two
lobes are highly asymmetric in intensity. The north-western component
has a hotspot which is brighter than the peak of emission at the outer
extremeties of the southern lobe by a factor of $\sim$10 (e.g. Bridle, Fomalont 
\& Cornwell 1981; Beswick et al.2004). The prominent central source 
has a steep radio spectrum, and when observed with high angular resolution
resembles a compact double-lobed source with multiple components and a
flat-spectrum radio core (Akujor et al. 1996; Beswick et al. 2004,
and references therein; Giovannini et al. 2005). The projected linear 
separation of the two prominent peaks on opposite sides of the nucleus
is $\sim$1.7 kpc, and has been interpreted to represent a more recent 
cycle of AGN activity, reminiscent of 3C236. 
However, 3C293 has a relatively small overall projected linear size compared 
with most DDRGs which are over approximately a Mpc in size. Also, the inner
double of 3C293 has diffuse extended emission beyond the two prominent peaks;
the total extent of the inner source including the extended emission is
$\sim$4.2 kpc. Akujor et al. (1996) find the diffuse extended emission to have
significantly steeper spectra than the inner components and have considered the
possibility that this might represent yet another cycle of activity. Similar
resolution data over a larger frequency range would be useful to further explore 
this possibility. 

In this paper, we present the results of low-frequency observations made 
with the Giant Metrewave Radio Telescope (GMRT) at 154, 240 and 614 MHz,
as well as Very Large Array (VLA) observations at 4860 MHz made from archival
data. These observations do not resolve the different components of the central
source seen in the high-resolution images. We have determined the spectra of the outer 
and inner lobes over a large frequency range, estimated their spectral ages and 
discuss the constraints on time scale of episodic activity.
The observations and data reduction are described in Section~2. The observational 
results, including the radio maps and spectra are presented in Section~3.  
The results are presented in Section~4, while the concluding remarks are given 
in Section~5.

\section{Observations and data reduction}
The analysis presented in this paper is based on radio observations
made with the GMRT, and as well as on VLA archival data.
The observing log for both the GMRT and VLA observations is 
listed in Table~1. 

\subsection{GMRT observations}
The GMRT observations were made in the standard manner, with  each observation
of the target-source interspersed with observations of the phase calibrator.
3C286 was observed as the flux density and bandpass calibrator, and all flux
densities are on the Baars et al. (1977) scale using the latest VLA values.
At each frequency the source was observed in a full-synthesis run of 
approximately 9 hours including calibration overheads.
The data were calibrated and reduced in the standard way using the NRAO {\tt AIPS} 
package.  Several rounds of self calibration were done to improve the quality 
of the images. 

\begin{table}
\caption{Observing log. Columns 1 and 2 show the name of the
telescope, and the array configuration for the VLA observations;
columns 3 and 4 show the frequency and bandwidth used in making the images;
column 5: the primary beamwidth in arcmin; 
column 6: dates of the observations.}
\begin{tabular}{l c r c c c }
\hline
Teles-    & Array  & Obs.   & Band- & Primary &  Obs. Date  \\
cope      & Conf.  & Freq.  & width &  beam   &             \\
          &        &(MHz)   &(MHz)  &(arcmin) &             \\
 (1)      &  (2)   & (3)    & (4)   &  (5)    &    (6)      \\
\hline
GMRT      &        & 154    &   6   &  186    & 2009 Jan 27   \\
GMRT      &        & 240    &   6   &  114    & 2008 Dec 22   \\
GMRT      &        & 614    &  16   &   43    & 2008 Dec 22   \\
VLA$^a$   &    C   & 4860   &  50   &   9     & 1984 Oct 04  \\
\hline
\end{tabular}

$^a$: VLA archival data.
\end{table}

\section{Observational results}
The observational results on 3C293 are described here. A catalogue of sources
detected at 154 MHz within 2.2$^\circ$ of the phase centre, i.e. the 
position of 3C293, was made  (HPBW$\sim$3.1$^\circ$).  Those with angular sizes 
larger than 45 arcsec were compared with the NRAO VLA Sky Survey 
(NVSS; Condon et al. 1998) images to examine evidence 
of any steep-spectrum relic emission not detected in the NVSS images at 1400 MHz.
No such relic emission was seen in the 154-MHz images, consistent with earlier
studies (e.g. Sirothia et al. 2009a), while a possible relic
and a couple of extended sources, all seen in the NVSS images, are 
described briefly in the Appendix. The catalogue of sources detected at 154 MHz 
with an angular resolution of 15$\times$12.5 arcsec$^2$ along
PA 73$^\circ$ is available electronically as on-line supplementary material. 
A total of 320 sources, including 3C293, which have a peak flux density 
$\geq$7-$\sigma$ have been listed. The value of $\sigma$, the primary beam corrected
local rms noise, varies from 8.6 mJy beam$^{-1}$ near 3C293 to typical values of
3$-$4 mJy beam$^{-1}$ in regions without a strong source. All but three of the 
weaker sources, namely, J1352+3059, J1352+3039 and J1357+3111, are seen in the NVSS 
images at 1400 MHz. Using 5 times the rms noise in the NVSS images yields spectral
indices steeper than $\sim$1.2 for these three sources.
  
\subsection{Overall structure of 3C293}
The observational parameters and the flux densities estimated from these images are
presented in Table 2. These have been estimated from the images made by tapering and 
weighting the data and convolving the images to match the resolution of the GMRT 
image at 154 MHz. The flux densitites have been estimated by specifying a similar 
area around the components at the different frequencies. We have examined the 
variation in flux density while varying the size of the area around the component.
Considering these fluctuations as well as calibration errors, the error in the flux 
density has been estimated to be approximately 15 per cent at 150 and 240 MHz, 7 per 
cent at 610 MHz and 5 per cent at 4860 MHz.

The source is quite asymmetric in both brightness and location of the outer components.
The ratio of separations from the core of the brighter north-western component to the
weaker south-eastern one is $\sim$0.6. The peak brightness of the north-western
component is larger by a factor of $\gapp$10, while the ratio of the total flux densities 
lies between a factor of $\sim$5 and 7, depending on the frequency of observations 
(see Table 2). The 
central source is also highly asymmetric in location, with the ratio of separations from
the core of the eastern component to the western one being $\sim$3. However, the flux
densities are more symmetric with the peak brightness ratio being within a factor of $\sim$2,
and the ratio of total flux densities varying between $\sim$1.2 and 1.6 (cf. Akujor et al.
1996; Beswick et al. 2004). Such asymmetries are often seen in compact steep-spectrum
radio sources (cf. Saikia et al. 1995), and the asymmetries observed in 3C293 are likely 
to be due to a combination of the effects of both orientation and relativistic motion 
(cf. Jeyakumar et al. 2005).  
  
\begin{table}
\caption{The observational parameters and flux densities.
Column 1: frequency  of observations in MHz, with the letter G or V representing
either GMRT or VLA observations; columns 2$-$4: the major and minor axes of the restoring
beam in arcsec and its PA in degrees; column 5: the rms noise in units of
mJy beam$^{-1}$; column 6: component designation, where Cent refers to the central source
including the radio core; columns 7 and 8: the peak and total flux densities of the source. }
\begin{tabular}{l rrr c l r r}
\hline
Freq.        & \multicolumn{3}{c}{Beam size}   & rms &  Cmp. & S$_p$  & S$_t$    \\
(MHz)        & ($^{\prime\prime}$) & ($^{\prime\prime}$) & ($^\circ$) &    (mJy) &  & (mJy)   &  (mJy)  \\
             &         &       &       & /b)    &     & /b)   &             \\ 
(1)          & (2)     & (3)   & (4)   & (5)    & (6) &  (7)   & (8)      \\ 
\hline
  G154       & 23.2    & 21.6  &   79  &    8.6 & NW  &  2198 &  5491   \\
             &         &       &       &        & Cent& 15216 & 15570    \\
             &         &       &       &        & SE  &   215 &  1022     \\
             &         &       &       &        & Tot & 15216 & 22288     \\

  G240       & 23.2    & 21.6  &   79  &   11.7 & NW  &  1401 &  3244       \\
             &         &       &       &        & Cent& 10825 & 11041     \\
             &         &       &       &        & SE  &   127 &   577        \\
             &         &       &       &        & Tot & 10825 & 15142     \\

  G614       & 23.2    & 21.6  &   79  &   3.0  & NW  &   754 &  1509      \\
             &         &       &       &        & Cent&  6421 &  6446       \\
             &         &       &       &        & SE  &    61 &   300       \\
             &         &       &       &        & Tot &  6421 &  8282     \\

  V4860      & 23.2    & 21.6  &   79  &   0.67 & NW  &  178  &  305       \\
             &         &       &       &        & Cent& 1556  & 1553        \\
             &         &       &       &        & SE  &   10  &   42        \\
             &         &       &       &        & Tot & 1556  & 1909      \\
\hline
\end{tabular}
\end{table}

%%%%%%%%%%%%%%%%%%%%%

%%%%%
\begin{figure}
\hbox{
 \psfig{file=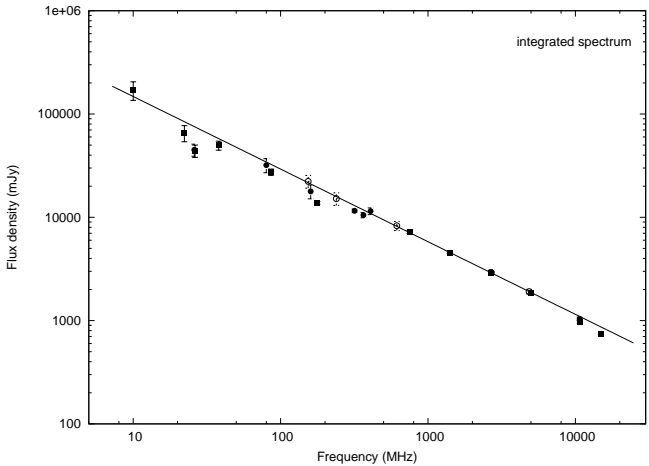,width=3.5in,angle=-90}
     }
\caption[]{The integrated spectrum of 3C293 using the
measurements from Laing \& Peacock (1980),
K\"uhr et al. (1981) and our measurements. These are shown
as filled squares,  and filled circles and open circles 
respectively.}
\end{figure}
%%%%%%

%%%%%
\begin{figure}
\vbox{
 \psfig{file=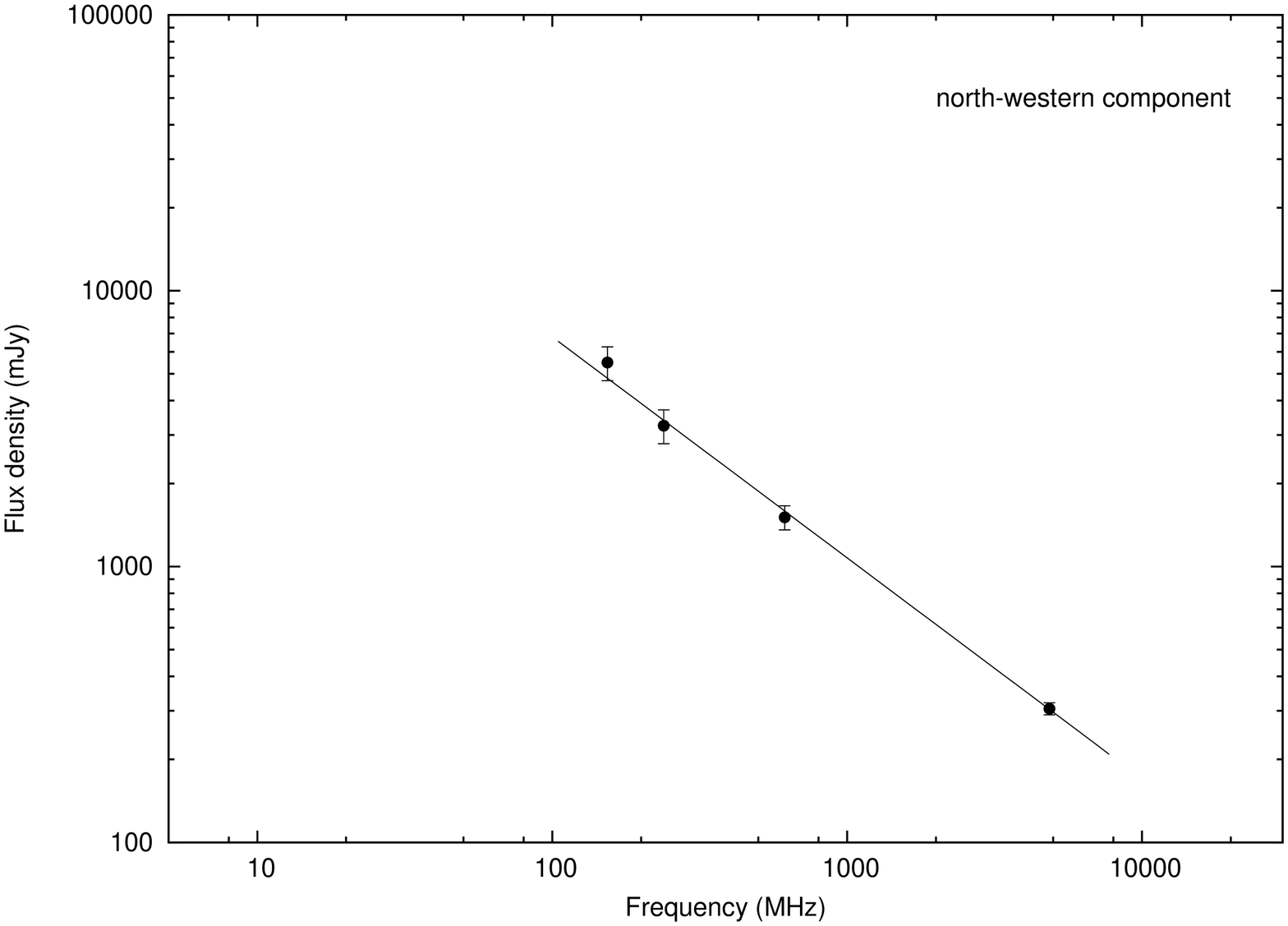,width=3.5in,angle=-90}
 \psfig{file=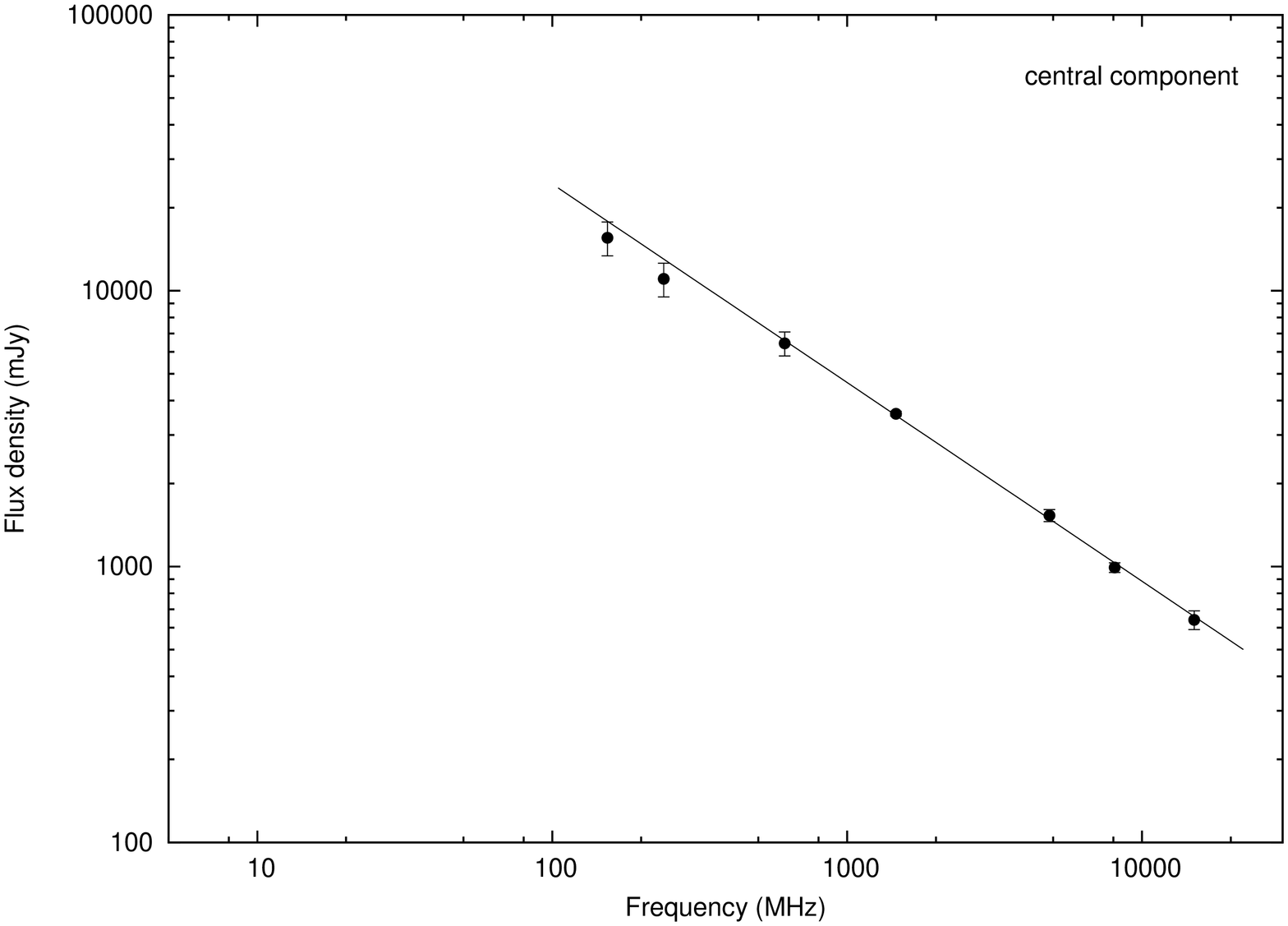,width=3.5in,angle=-90}
 \psfig{file=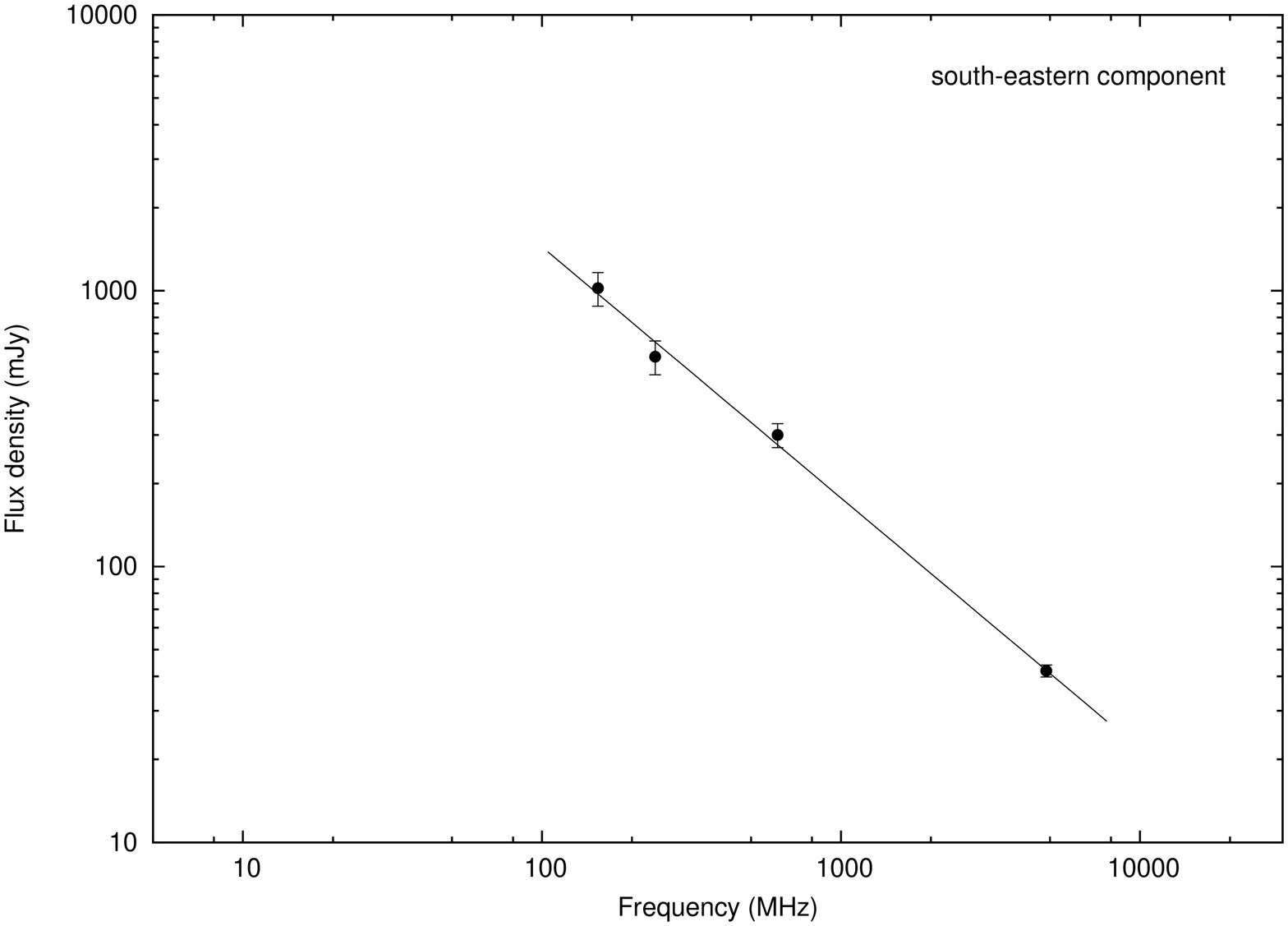,width=3.5in,angle=-90}
     }
\caption[]{The spectra of the north-western (upper),
central (middle) and south-eastern (lower) components
respectively. For the central component, the core flux
density has been subtracted for measurements above
$\sim$1400 MHz, as described in the text. 
           }
\end{figure}
%%%%%%

\subsection{Spectra}
The integrated spectrum of 3C293 using the data from Laing \& Peacock (1980)
and K\"{u}hr et al. (1981), as well as from our measurements is shown in Fig. 2. 
All the flux densities are consistent with the scale of Baars et al.  (1977). 
Excluding the points below 50 MHz which have large uncertainties and the 
measurement at 178 MHz which is significantly lower than expected from the
measurements at 154 and 160 MHz, the spectral index, $\alpha$ (S$\propto\nu^{-\alpha}$), 
of the source is 0.71$\pm$0.01. Using our measurements alone yields the same
value for the spectral index.

In Fig. 3  we present the spectra of the north-western, central and south-eastern 
components using our measurements. For the central component we have also used the
measurements made by Bridle et al. (1981) after subtracting the flux density of the
flat-spectrum core component at frequencies of $\sim$1400 MHz and above using the
values listed by Bridle et al. (1981), Akujor et al. (1996) and Beswick et al. (2004).  
The core is relatively weak, even at $\sim$15000 MHz, and has made no significant
difference to the spectra. The flux densities of the extended lobes have not been
listed by Bridle et al. (1981). The flux densities of all the three components are consistent
with straight spectra, the spectral indices being 0.72$\pm$0.02 for the central 
component, and 0.80$\pm$0.02 and 0.91$\pm$0.03 for the north-western and south-eastern
components respectively. The spectral index of the central component, which contributes
over 70 per cent of the total flux density of the source, is consistent with the 
integrated spectral index.  

\section{Discussion}
\subsection{Spectral ages} 
The spectra of the outer components using the total flux densities between 150 
and 4860 MHz and that of the inner double using measurements between 150 and 15000 MHz 
are consistent with a single power-law.  We have fitted the spectrum of the inner double 
after subtracting the core flux density for the
Jaffe \& Perola (1973; JP), Kardashev-Pacholczyk (KP, Kardashev 1962; Pacholczyk 1970) and
the continuous injection (CI, Pacholczyk 1970) models using the {\tt SYNAGE} package (Murgia et al. 1999).  
The break frequency obtained from these fits are rather large ($\gapp$7$\times$10$^5$ GHz)
and have huge uncertainties because the spectra are practically straight. We have adopted
conservative lower limits to the break frequency to be the highest frequency 
of our observations.

In order to estimate the spectral ages, we have to estimate the magnetic field strength,
which was done in a similar way to that of 3C46 and 3C452 (Nandi et al. 2010). 
The magnetic field strength has been estimated by integrating the spectrum from
a frequency corresponding to a minimum Lorentz factor,
$\gamma_{\rm min}$ $\sim$10 for the relativistic electrons to an upper limit of
100 GHz, which corresponds to a Lorentz factor ranging
from a few times 10$^4$ to 10$^5$ depending on the estimated
magnetic field strength (see Hardcastle et al. 2004; Croston
et al. 2005; Konar et al. 2008, 2009). It has also been assumed that the filling factors of 
the lobes are unity, and the
energetically dominant particles are the radiating particles
only, neglecting the contribution of the protons. 
A cylindrical geometry has been assumed for both the lobes, and the central source,
and their sizes have been estimated more reliably from the availabe higher-resolution
images (e.g. Bridle et al. 1981; Akujor et al. 1996). The deconvolved size of the central
source estimated from our observations is consistent with that estimated from the
higher-resloution image of Akujor et al. (1996).   

The equipartition magnetic field estimate for the inner double
is 16.92$\pm$1.67 nT, indicating that for a conservative break frequency $\gapp$16 GHz,
the inferred spectral age is $\lapp$0.18 Myr. 
For the extended lobes, where reliable measurements of the total flux density are 
available up to $\sim$5 GHz, the magnetic field strengths are 1.12$\pm$0.11 and 0.88$\pm$0.09 nT 
for the north-western and south-eastern lobes respectively, while the corresponding spectral ages 
are $\lapp$16.9 and 23.0 Myr respectively for a break frequency $\gapp$5 GHz. For a break 
frequency of $\gapp$100 GHz, as was adopted by Nandi et al. (2010), the
spectral ages are $\lapp$0.07, 3.77 and 5.13 Myr respectively for the central source, and
the north-western and south-eastern components respectively. Machalski et al. (2007) have 
examined the dynamical ages of FRII radio 
sources and find that these agree with the spectral ages for objects less than 10 Myr, while
for a sample of giant radio galaxies the ratio of dynamical age to the spectral age of the 
lobes lies between $\sim$1 and 5 (Machalski, Jamrozy \& Saikia 2009). Lower limits to the
dynamical age may be estimated by assuming the velocity of advancement to be c, and an inclination of
the source axis to the line of sight to be $\sim$45$^\circ$. This yields ages of $\gapp$0.3 Myr
for the outer north-western hot-spot, and $\gapp$0.006 Myr for the more distant eastern hot-spot
of the central source in 3C293. 

\subsection{Recurrent activity time scale} 
In addition to the age estimates, one can also constrain the time scale of interruption of
jet activity from the presence of a hot-spot in the north-western lobe of 3C293.
Hot-spots in the outer lobes have been seen in other sources with evidence of episodic
jet activity, although the outer doubles are  often diffuse as for example
in J1453+3308 (Schoenmakers et al. 2000a; Konar et al. 2006).
Examples of sources with hot-spots are the northern lobe of B1834+620 
(Schoenmakers et al. 2000b), and the western outer lobe of 4C02.27 
(Jamrozy, Saikia \& Konar 2009). The hot-spots have typical sizes of $\lapp$10 kpc
(e.g. Jeyakumar \& Saikia 2000),
and are expected to fade in $\sim$10$^4$ to 10$^5$ yr after the energy supply has
been cut off (e.g. Kaiser, Schoenmakers \& Rottgering 2000). This is usually smaller
than the time it takes for the material to reach the hot-spots from the radio core,
and hence it is reasonable to assume that the hot-spot fades soon after the last jet
material has passed through it. 

For 3C293, the fraction of emission from the core, f$_c$, at an emitted frequency of 8 GHz, which 
is often used as a statistical indicator of the orientation of the jet axis to the line
of sight, is $\sim$0.017. The corresponding median value of f$_c$ for galaxies is $\sim$0.002
while for quasars it is $\sim$0.05 (e.g. Saikia \& Kulkarni 1994), suggesting that the orientation 
of the nuclear jet is close to the dividing line between radio galaxies and quasars. We adopt an
orientation angle of $\sim$45$^\circ$ to the line of sight. The hot-spot in the north-western
lobe implies that it still receives jet material. For an inclination angle, $\phi$$\sim$45$^\circ$
and a jet velocity of c, the travel time from the core to the hot-spot is $\sim$0.35 Myr,
However, this will be affected by light-travel time effects due to the orientation of 
the source axis. The high brightness asymmetry and the observed levels of polarization 
(cf. Bridle et al. 1981) suggest that the north-western lobe is approaching us, indicating
that the observed time difference between the ejection of the last material and its
arrival at the approaching hot-spot is $\sim$0.1 Myr. To be able to see the hot-spot as well
as the inner structure, the interrruption of jet activity must be less than $\sim$0.1 Myr, within 
which period of time the inner double must also form. The estimated spectral age of the inner
double assuming a break frequency of $\sim$100 GHz is consistent with this scenario. 

\section{Concluding remarks}
Estimation of the time scales of recurrent activity is important to constrain models 
of such activity (e.g. Kaiser, Schoemakers \& R\"ottgering 2000; Brocksopp et al. 2011), 
as well as understand possible effects on the host galaxy and its evolution.  
GMRT and VLA observations of this highly misaligned DDRG have shown that 
the spectral age of the inner double is likely to be $\lapp$0.1 Myr, which
is similar to the time scale of interruption of jet activity. The overall
linear size of $\sim$190 kpc, is much smaller than most DDRGs which are
known to be over  a Mpc in size. For example in the well-studied DDRG 
J1453+3308, the dynamical and spectral ages of the diffuse
outer lobes of are $\sim$215 and 50 Myr, while that of the
inner double is only $\sim$2 Myr, suggesting a much longer time-scale
of interruption (Kaiser et al. 2000; Konar et al. 2006). Reliable
identification of signs of episodic activity in sources of different sizes is
required to explore the entire range of time scales of episodic activity. For example
in the compact steep-spectrum source CTA~21, a compact double-lobed structure with
a size of $\sim$12 mas (Jones 1984; Kellermann et al. 1998),  and diffuse emission
on scales extending to $\sim$300 mas (Dallacasa et al. 1995), led Salter et al. (2010)
to suggest that CTA~21 may be undergoing repeated cycles of activity. This is consistent
with the suggestion of (Reynolds \& Begelman 1997) that jet activity in compact sources
may be intermittent on time scales of 10$^4$ to 10$^5$ yr. Objects such as 3C293 help
us explore the implications of the intermediate range of time scales of episodic nuclear 
activity. For example the model by Kaiser et al. (2000) where long time-scales of $\sim$10$^7$
yr is required for the dispersion of the clouds of entrained material would not be applicable
here. Here, given the small size of the inner double, the hotspots are likely to have formed
by interaction of the jets with the interstellar medium of the host galaxy.

\section*{Acknowledgments} 
We thank an anonymous reviewer for several helpful comments, 
the staffs of GMRT and  VLA for their help with the
observations, and Sandeep Sirothia for generating the catalogue of sources using 
his software. SN thanks NCRA for hospitality, and DST, Government of India  for
financial support via grant number SR/S2/HEP-17/2005. The GMRT is a national facility 
operated by the National Centre for Radio Astrophysics of the Tata Institute of Fundamental 
Research.  The National Radio Astronomy Observatory  is a facility of the National Science 
Foundation operated under co-operative agreement by Associated Universities Inc. 
This research has made use of the NASA/IPAC extragalactic database (NED) which is operated
by the Jet Propulsion Laboratory, Caltech, under contract with the National Aeronautics
and Space Administration.  We thank Matteo Murgia for access to the {\tt SYNAGE} software.

{}

\section*{Appendix}
The images of three of the extended sources detected at 154 MHz are shown
here, along with an extract from the catalogue of 320 sources detected at this
frequency in the image with an angular resolution of 15$\times$12.5 arcsec$^2$ along 
PA 73$^\circ$. The catalogue consists of sources with a peak flux density $\geq$7 
times the local rms noise value.  The full table is available as on-line supplementary 
material.  Amongst the extended sources detected at this frequency, 
Fig. 4 shows a wide-angle tailed (WAT) source associated with a galaxy
at a redshift of 0.061. This was detected $\sim$2.5$^{\circ}$ north of the
phase centre. A lower resolution image of the source shows diffuse emission
towards the south-east which has been estimated to have a spectral index of
$\sim$1.02  between 154 and 1400 MHz using the NVSS image, while the WAT has
a spectral index of $\sim$0.91. The diffuse blob has a tail extending approximately
in the direction of the WAT host galaxy, and is likely to be related 
to the WAT source. There is no known optical counterpart for this source.

\begin{figure}
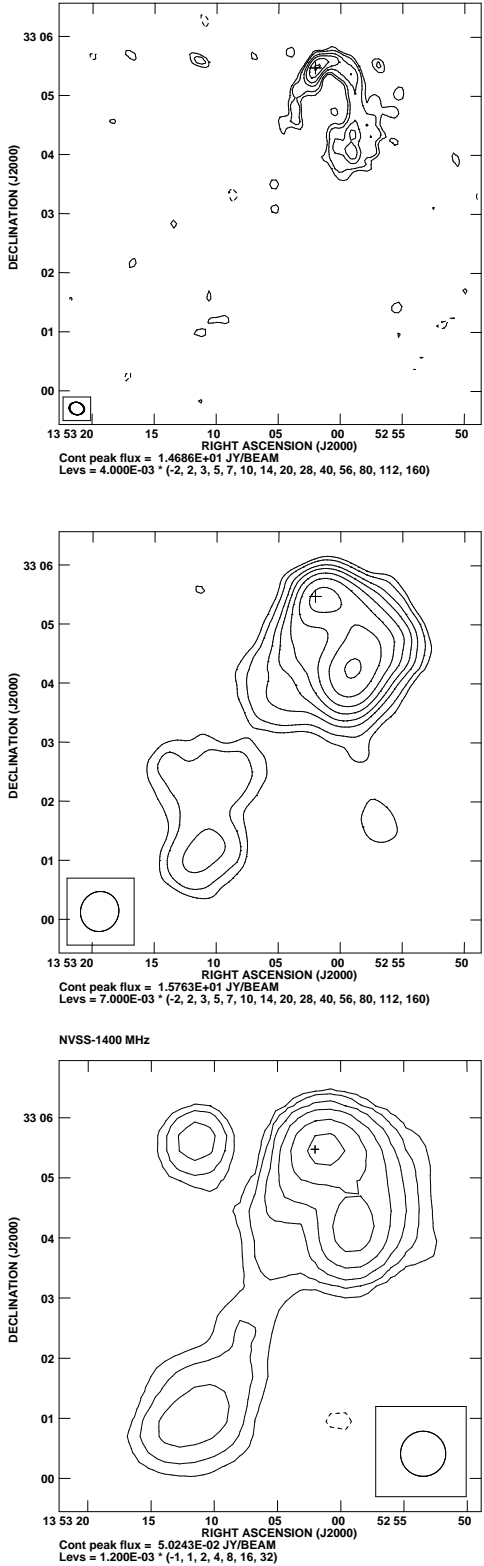

\vbox{
\psfig{figure=relic_highres.ps,width=6.6cm,angle=0} 
\psfig{figure=relic_lowres.ps,width=6.6cm,angle=0} 
\psfig{figure=wat.nvss.ps,width=6.6cm,angle=0} 
     }
\caption{A WAT source associated a galaxy at a redshift of 0.061 along with
  evidence of diffuse emission. The upper panel shows the higher resolution
  GMRT image at 154 MHz with a restoring beam of 
  15$\times$12.5 arcsec$^2$ along PA 73$^\circ$, while in the middle panel
  the restoring beam is 41$\times$39 arcsec$^2$ along PA 159$^\circ$. The lower
  panel shows the NVSS image at 1400 MHz.  Crosses mark 
  the position of the optical galaxy.} 
\label{fig:relic}
\end{figure}

Approximately 2.5$^{\circ}$ to the south-east
of the phase centre we detect the large radio galaxy J1400+3019 (Fig. 5, upper panel), 
which is associated with a galaxy at a redshift of 0.206, 
and has an overall angular size of 649 arcsec, indicating a projected linear
size of 2170 kpc (Parma et al. 1996; Ishwara-Chandra \& Saikia 1999). The spectral
index of this source between 154 and 1400 MHz is $\sim$1.01.

Approximately 30 arcmin towards the west of 3C293, the 154-MHz image shows
an FRI galaxy, J1356+3126 (Fig. 5, lower panel), which resembles the morphology of 
3C31. There are three galaxies in the vicinity of the centroid, of which the
one closest to the peak, SDSS J135620.77+312627.3  with a redshift of 0.151 is
likely to be the identification. The nearby companion SDSS J135621.36+312631.7 
is only $\sim$8.7 arcsec towards the north-east. Although its redshift has not
been measured, its separation would be $\sim$23 kpc at a redshift of 0.151 and
could be a companion galaxy. It may be relevant to note that NGC383, the galaxy 
associated with 3C31 has a nearby companion and is a member of a chain of galaxies
Arp 331 (e.g. Strom et al. 1983).  The overall angular size of the radio source 
J1356+3126 is $\sim$120 arcsec which corresponds to a projected linear 
size of $\sim$310 kpc. Its spectral index is 0.6 between 154 and 1400 MHz.

\begin{figure}
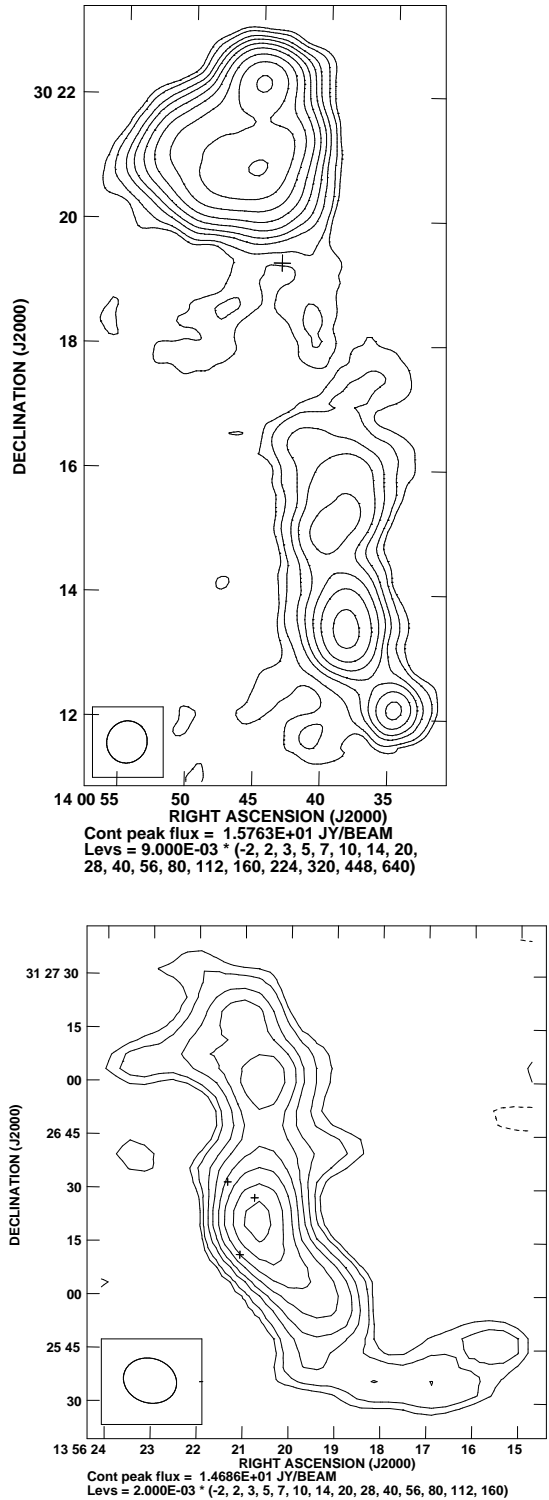

\vbox{
\psfig{figure=fr2.sdss.ps,width=6.2cm,angle=0} 
\psfig{figure=fr1.3opt.ps,width=7.5cm,angle=0} 
     }
\caption{A known giant radio galaxy with an FRII-type structure (upper panel)
  and an FRI-type galaxy resembling the morphology of 3C 31 (lower panel).
  Crosses in the images mark the positions of the optical galaxies.
  The restoring beam for the giant radio galaxy, J1400+3019, is 
  41$\times$39 arcsec$^2$ along a  PA=159$^\circ$, while  for the FRI 
  galaxy, J1356+3126, it is 15$\times$12.5 arcsec$^2$ along  PA=73$^\circ$.
        }
\label{fig:etc}
\end{figure}	

\begin{table}
\caption{The first 15 sources from the list of sources detected at 154 MHz 
within 2.2$^\circ$ of the phase centre (HPBW$\sim$3.1$^\circ$). 
Sources with a peak flux density
$\geq$7-$\sigma$, where  $\sigma$ is the primary beam corrected local
rms noise in units of mJy beam$^{-1}$ have been listed. There are a total 
of 320 sources.  Columns 1 and 2: The right
ascension (h:m:s) and declination (d:m:s) in J2000 co-ordinates;
columns 3 and 4: the peak and integrated flux densities in units of mJy
beam$^{-1}$ and mJy respectively estimated from the primary beam 
corrected image; column 5: the angular size of the
source in arcsec where U denotes an unresolved source. The flux densities and
source sizes have been estimated as described in Sirothia et al. (2009b). }

\begin{tabular}{c c r r r}
RA (h:m:s)   & Dec (d:m:s)   & S$_{peak}$ & S$_{int}$ & Size \\
             &               & mJy/beam   &   mJy     & $^{\prime\prime}$ \\           
\hline
13:42:08.04 & 31:19:32.6 & 172.1 & 330.4 & 30.7 \\
13:42:31.58 & 32:03:29.1 &  36.6 &  36.6 & U \\
13:42:51.04 & 31:37:56.4 &  67.2 &  67.2 & U \\
13:42:51.39 & 31:11:04.6 & 204.3 & 518.0 & 95.4 \\
13:43:03.02 & 31:43:50.0 & 2341.9 & 3961.1 & 59.5 \\
13:43:07.66 & 31:27:50.3 & 319.2 & 497.4 & 44.2 \\
13:43:08.30 & 30:31:10.7 &  34.0 &  34.0 & U \\
13:43:13.10 & 32:08:01.0 &  33.0 &  33.0 & U \\
13:43:28.67 & 31:16:30.0 &  61.2 &  61.2 & U \\
13:43:53.47 & 32:43:28.5 &  36.6 &  36.6 & U \\
13:44:00.76 & 32:11:27.3 & 148.1 & 148.1 & U \\
13:44:26.84 & 31:34:51.7 & 424.3 & 934.2 & 39.5 \\
13:44:36.79 & 31:04:25.1 & 179.3 & 221.8 & 12.6 \\
13:44:39.72 & 32:01:23.1 & 1519.7 & 2003.7 & 61.6 \\
13:44:54.55 & 31:51:43.5 &  34.0 &  34.0 & U \\

\hline
\end{tabular}
\end{table}

%%%%%%%%%%%%%%%%%%%%%%%%%%%%%%%%%%%%%%%%%%%%%%%%%
\clearpage

\begin{table}
\caption{Sources within 2.2$^\circ$ of the phase centre 
  of the GMRT observations at 154 MHz 
    (HPBW$\sim$3.1$^\circ$). All sources with a peak flux density
  $\geq$7-$\sigma$, where  $\sigma$ is the primary beam corrected
  local rms noise in units of mJy beam$^{-1}$ are listed. 
 The values of
    $\sigma$ range from 8.6 mJy beam$^{-1}$ near 3C293 to typical
    values of 3$-$4 mJy beam$^{-1}$ in regions without strong sources. There
  are a total of 320 sources. All but three of the weaker sources
    (J1352+3059, J1352+3039 and J1357+3111) are seen in the NVSS
    images at 1400 MHz. Using 5 times the rms noise in the NVSS images
    yields spectral indices steeper than $\sim$1.2 for these three
    sources. Columns 1 and 2: The right ascension (h:m:s) and
  declination (d:m:s) in J2000 co-ordinates; columns 3 and 4: the peak
  and integrated flux densities in units of mJy beam$^{-1}$ and mJy
  respectively; column 5: the angular size of the source in arcsec
  where U denotes an unresolved source. The flux densities and source
  sizes have been estimated as described in Sirothia et al. (2009b).  }

\begin{tabular}{ l l c  r r}
\hline \hline
RA (h:m:s) & Dec (d:m:s) & S$_{\rm {peak}}$ & S$_{\rm {int}}$ & Size \\
           &             &  mJy beam$^{-1}$ &    mJy          & $^{\prime\prime}$  \\
   (1)     &    (2)      &        (3)       &       (4)       &  (5) \\
\hline
13:42:08.04 & 31:19:32.6 & 172.1 & 330.4 & 30.7 \\ 
13:42:31.58 & 32:03:29.1 &  36.6 &  36.6 & U \\ 
13:42:51.04 & 31:37:56.4 &  67.2 &  67.2 & U \\ 
13:42:51.39 & 31:11:04.6 & 204.3 & 518.0 & 95.4 \\ 
13:43:03.02 & 31:43:50.0 & 2341.9 & 3961.1 & 59.5 \\ 
13:43:07.66 & 31:27:50.3 & 319.2 & 497.4 & 44.2 \\ 
13:43:08.30 & 30:31:10.7 &  34.0 &  34.0 & U \\ 
13:43:13.10 & 32:08:01.0 &  33.0 &  33.0 & U \\ 
13:43:28.67 & 31:16:30.0 &  61.2 &  61.2 & U \\ 
13:43:53.47 & 32:43:28.5 &  36.6 &  36.6 & U \\ 
13:44:00.76 & 32:11:27.3 & 148.1 & 148.1 & U \\ 
13:44:26.84 & 31:34:51.7 & 424.3 & 934.2 & 39.5 \\ 
13:44:36.79 & 31:04:25.1 & 179.3 & 221.8 & 12.6 \\ 
13:44:39.72 & 32:01:23.1 & 1519.7 & 2003.7 & 61.6 \\ 
13:44:54.55 & 31:51:43.5 &  34.0 &  34.0 & U \\ 
13:44:58.49 & 30:15:53.1 & 334.6 & 699.0 & 85.5 \\ 
13:45:10.70 & 32:10:53.7 & 404.8 & 674.3 & 23.6 \\ 
13:45:29.69 & 32:27:06.8 &  44.6 &  44.6 & U \\ 
13:45:38.98 & 30:59:15.5 &  41.2 &  41.2 & U \\ 
13:45:41.18 & 30:08:35.9 &  36.0 &  36.0 & U \\ 
13:45:41.63 & 31:24:05.6 & 181.1 & 336.2 & 31.1 \\ 
13:45:43.02 & 31:33:41.6 & 102.2 & 125.4 & 14.6 \\ 
13:45:48.10 & 32:32:28.0 &  35.6 &  35.6 & U \\ 
13:45:49.14 & 31:41:19.9 &  32.2 &  32.2 & U \\ 
13:45:49.56 & 32:07:57.9 &  34.4 &  34.4 & U \\ 

\end{tabular}
\end{table}

\begin{table}
\begin{tabular}{l l c r r}
\hline
RA (h:m:s) & Dec (d:m:s) & S$_{\rm {peak}}$ & S$_{\rm {int}}$ & Size \\
\hline

13:45:50.50 & 31:03:25.4 &  49.8 &  49.8 & U \\ 
13:45:54.80 & 33:07:55.7 &  56.3 &  56.3 & U \\ 
13:45:57.01 & 30:18:19.9 &  41.3 &  41.3 & U \\ 
13:45:57.88 & 32:47:26.6 &  33.1 &  33.1 & U \\ 
13:45:59.25 & 32:29:25.1 &  47.4 &  47.4 & U \\ 
13:46:12.01 & 30:17:01.9 &  66.2 &  66.2 & U \\ 
13:46:12.15 & 32:34:18.2 & 230.8 & 326.3 & 33.6 \\ 
13:46:17.27 & 30:06:47.6 & 123.9 & 271.8 & 41.5 \\ 
13:46:21.31 & 32:49:00.9 & 303.2 & 495.1 & 30.6 \\ 
13:46:27.07 & 30:20:58.4 &  54.6 &  54.6 & U \\ 
13:46:29.65 & 32:51:19.1 &  86.3 & 165.7 & 28.0 \\ 
13:46:34.13 & 31:29:40.5 &  81.6 & 159.7 & 18.3 \\ 
13:46:40.04 & 32:34:17.8 &  27.7 &  27.7 & U \\ 
13:46:48.00 & 30:05:01.9 & 153.8 & 153.8 & U \\ 
13:46:51.53 & 29:45:34.2 &  53.5 &  53.5 & U \\ 
13:46:54.61 & 32:19:03.4 &  88.1 &  88.1 & U \\ 
13:46:57.66 & 31:16:53.8 & 165.1 & 165.1 & U \\ 
13:46:58.72 & 30:03:28.5 &  64.1 &  64.1 & U \\ 
13:47:01.80 & 31:09:12.5 &  46.4 & 354.3 & 98.5 \\ 
13:47:15.55 & 30:33:17.0 & 582.1 & 783.1 & 78.4 \\ 
13:47:16.33 & 32:17:16.7 &  45.0 &  45.0 & U \\ 
13:47:17.65 & 31:24:42.4 &  59.5 & 163.8 & 64.7 \\ 
13:47:19.13 & 32:21:46.6 & 103.7 & 103.7 & U \\ 
13:47:19.87 & 31:21:19.8 &  69.4 &  69.4 & U \\ 
13:47:26.99 & 31:17:23.9 &  84.5 &  84.5 & U \\ 
13:47:31.19 & 30:45:52.4 & 170.8 & 221.4 & 38.4 \\ 
13:47:31.71 & 32:03:45.3 &  50.4 &  50.4 & U \\ 
13:47:34.69 & 31:28:15.0 &  21.5 &  21.5 & U \\ 
13:47:37.60 & 31:59:15.5 &  57.4 &  57.4 & U \\ 
13:47:40.17 & 32:12:01.5 &  29.0 &  29.0 & U \\ 
13:47:47.99 & 32:58:18.9 & 405.0 & 1112.1 & 74.1 \\ 
13:47:52.54 & 33:22:20.3 &  70.9 &  70.9 & U \\ 
13:47:54.69 & 32:06:13.7 & 128.9 & 156.0 & 13.4 \\ 
13:47:55.14 & 32:00:42.7 & 110.6 & 163.0 & 72.0 \\ 
13:47:58.59 & 33:00:33.4 &  52.8 &  52.8 & U \\ 
13:48:00.52 & 33:03:57.0 &  72.2 &  72.2 & U \\ 
13:48:01.86 & 31:05:48.3 &  38.4 &  38.4 & U \\ 
13:48:03.04 & 32:26:07.2 &  31.7 &  31.7 & U \\ 
13:48:03.34 & 30:10:18.0 & 231.9 & 543.9 & 56.6 \\ 
13:48:04.30 & 30:52:30.1 &  38.3 &  63.4 & 27.3 \\ 
13:48:06.18 & 31:15:51.9 &  56.6 &  77.4 & 22.9 \\ 
13:48:06.82 & 33:23:28.9 &  65.8 &  65.8 & U \\ 
13:48:08.72 & 30:49:06.2 &  47.1 &  47.1 & U \\ 
13:48:11.57 & 30:37:56.8 &  42.6 &  42.6 & U \\ 
13:48:13.71 & 32:51:21.9 & 135.8 & 242.5 & 31.3 \\ 

\end{tabular}
\end{table}

\begin{table}
\begin{tabular}{l l  c  r r}
\hline
RA (h:m:s) & Dec (d:m:s) & S$_{\rm {peak}}$ & S$_{\rm {int}}$ & Size \\
\hline

13:48:17.11 & 31:03:10.1 &  34.3 &  70.4 & 48.5 \\ 
13:48:18.18 & 33:05:53.5 &  33.8 &  33.8 & U \\ 
13:48:20.32 & 30:20:02.9 & 117.3 & 320.7 & 134.8 \\ 
13:48:20.71 & 32:22:14.0 &  24.1 &  24.1 & U \\ 
13:48:30.41 & 32:54:58.8 &  46.9 &  46.9 & U \\ 
13:48:31.84 & 30:39:25.4 &  40.1 &  40.1 & U \\ 
13:48:42.13 & 31:33:04.7 & 163.5 & 191.0 & 26.7 \\ 
13:48:42.32 & 33:10:18.4 &  60.1 &  60.1 & U \\ 
13:48:48.02 & 29:39:18.9 & 2048.4 & 4951.0 & 44.3 \\ 
13:48:49.83 & 31:10:18.5 & 110.0 & 155.3 & 28.8 \\ 
13:48:51.08 & 32:02:28.4 &  63.9 & 121.7 & 48.6 \\ 
13:48:52.13 & 32:09:25.3 &  25.8 &  25.8 & U \\ 
13:48:55.10 & 31:57:32.4 & 394.0 & 502.4 & 41.4 \\ 
13:48:56.82 & 33:17:10.4 & 140.2 & 193.0 & 16.5 \\ 
13:48:57.33 & 29:28:53.7 &  45.7 &  45.7 & U \\ 
13:48:59.60 & 32:59:47.6 &  44.3 &  44.3 & U \\ 
13:49:03.00 & 31:40:09.4 &  34.4 &  34.4 & U \\ 
13:49:04.19 & 32:18:32.2 &  42.0 &  75.4 & 29.0 \\ 
13:49:08.61 & 30:51:50.2 &  43.5 &  43.5 & U \\ 
13:49:11.10 & 31:32:07.2 &  59.7 &  59.7 & U \\ 
13:49:11.48 & 31:40:31.6 &  32.8 &  32.8 & U \\ 
13:49:11.68 & 31:33:49.2 &  30.5 &  30.5 & U \\ 
13:49:14.00 & 33:03:16.9 &  71.7 &  71.7 & U \\ 
13:49:29.28 & 31:22:54.0 &  53.6 &  53.6 & U \\ 
13:49:34.12 & 30:04:34.0 &  39.9 &  39.9 & U \\ 
13:49:37.94 & 31:44:47.2 &  65.5 &  65.5 & U \\ 

13:49:39.76 & 30:08:48.8 &  76.0 &  76.0 & U \\ 
13:49:48.26 & 31:01:54.9 &  86.9 & 164.8 & 29.3 \\ 
13:49:50.51 & 29:35:37.5 &  74.0 &  74.0 & U \\ 
13:49:52.63 & 30:15:22.9 &  55.7 & 189.5 & 118.2 \\ 
13:49:58.19 & 31:04:13.1 &  23.8 &  23.8 & U \\ 
13:49:58.52 & 32:49:56.1 &  72.2 & 133.4 & 54.4 \\ 
13:50:00.33 & 30:04:15.2 & 124.7 & 197.0 & 16.6 \\ 
13:50:06.10 & 31:03:54.3 &  23.8 &  23.8 & U \\ 
13:50:11.22 & 29:42:54.0 &  49.2 &  49.2 & U \\ 
13:50:11.84 & 31:32:34.9 &  24.0 &  24.0 & U \\ 
13:50:11.97 & 30:22:44.8 & 705.9 & 1274.6 & 89.3 \\ 
13:50:16.38 & 32:36:59.4 &  36.2 &  36.2 & U \\ 
13:50:18.36 & 30:37:39.1 &  72.5 & 171.0 & 73.6 \\ 
13:50:19.39 & 30:22:39.8 &  72.7 &  72.7 & U \\ 
13:50:20.80 & 30:32:05.0 & 191.3 & 249.3 & 27.7 \\ 
13:50:31.47 & 30:57:24.9 & 191.2 & 282.2 & 72.7 \\ 
13:50:31.90 & 33:01:34.2 &  49.0 & 215.7 & 118.6 \\ 
13:50:32.50 & 30:10:47.9 &  39.7 &  39.7 & U \\ 
13:50:33.48 & 30:07:51.7 &  28.8 &  60.2 & 32.2 \\ 
13:50:35.43 & 30:40:17.3 &  53.8 &  53.8 & U \\ 
13:50:36.64 & 30:37:30.0 &  79.0 &  79.0 & U \\ 
13:50:39.43 & 29:31:05.9 &  39.9 &  39.9 & U \\ 
13:50:45.65 & 31:28:33.1 & 136.6 & 237.7 & 32.0 \\ 
13:50:46.59 & 30:13:25.7 & 101.8 & 101.8 & U \\ 
13:50:47.83 & 33:07:30.7 & 139.8 & 159.8 & 13.0 \\ 
13:50:48.29 & 33:12:12.3 & 626.9 & 1269.2 & 64.2 \\ 
13:50:49.26 & 33:05:31.0 &  70.4 &  70.4 & U \\ 
13:50:49.94 & 29:42:51.3 &  45.0 &  45.0 & U \\ 
13:50:50.05 & 30:09:12.3 & 244.4 & 319.4 & 26.6 \\ 
13:50:52.54 & 30:34:52.8 & 289.5 & 370.7 & 48.0 \\ 
13:50:56.90 & 33:31:46.3 & 179.0 & 443.3 & 32.7 \\ 
13:51:01.11 & 30:42:44.1 &  33.4 &  33.4 & U \\ 
13:51:01.97 & 32:34:35.9 &  51.7 &  51.7 & U \\ 
13:51:02.66 & 31:14:28.3 & 108.5 & 108.5 & U \\ 
13:51:03.14 & 30:53:56.1 & 338.7 & 678.4 & 57.1 \\ 
13:51:11.21 & 33:17:38.2 &  44.5 &  44.5 & U \\ 
13:51:12.55 & 33:21:51.7 & 127.0 & 249.6 & 37.0 \\ 

\end{tabular}
\end{table}

\begin{table}
\begin{tabular}{l l c r r}
\hline
RA (h:m:s) & Dec (d:m:s) & S$_{\rm {peak}}$ & S$_{\rm {int}}$ & Size \\
\hline

13:51:16.62 & 30:55:53.0 &  60.2 & 248.7 & 115.5 \\ 
13:51:20.97 & 33:07:09.9 & 197.3 & 197.3 & U \\ 
13:51:21.99 & 31:33:41.1 & 200.5 & 200.5 & U \\ 
13:51:23.30 & 30:07:37.8 &  61.1 &  61.1 & U \\ 
13:51:26.10 & 30:24:41.8 & 167.9 & 191.7 & 31.9 \\ 
13:51:27.82 & 31:04:01.9 & 121.6 & 147.8 & 35.2 \\ 
13:51:31.09 & 30:25:01.7 &  28.4 &  28.4 & U \\ 
13:51:35.18 & 29:33:53.7 & 108.4 & 201.0 & 41.3 \\ 
13:51:39.06 & 33:21:57.7 & 127.6 & 127.6 & U \\ 
13:51:41.54 & 31:04:43.4 &  40.4 &  40.4 & U \\ 
13:51:42.82 & 30:02:55.7 &  33.2 &  33.2 & U \\ 
13:51:47.14 & 32:06:14.7 &  36.8 &  36.8 & U \\ 
13:51:53.81 & 33:05:52.3 &  56.4 &  56.4 & U \\ 
13:52:05.31 & 32:52:27.8 &  32.8 &  32.8 & U \\ 
13:52:12.58 & 32:08:15.0 &  48.7 &  48.7 & U \\ 
13:52:15.83 & 32:29:09.6 &  37.6 &  37.6 & U \\ 
13:52:15.99 & 32:58:11.6 &  56.6 & 239.8 & 89.6 \\ 
13:52:16.76 & 31:26:55.2 & 15216.0 & 21060.6 & 131.5 \\ 
13:52:19.73 & 30:44:33.1 & 183.8 & 213.9 & 10.9 \\ 
13:52:21.29 & 32:49:05.0 &  31.8 &  31.8 & U \\ 
13:52:22.92 & 29:49:43.7 &  49.6 &  76.7 & 29.6 \\ 
13:52:23.49 & 31:25:45.8 & 213.5 & 895.2 & 74.5 \\ 
13:52:26.47 & 30:20:07.5 & 719.3 & 1096.8 & 60.9 \\ 
13:52:33.06 & 32:39:00.1 &  23.8 &  23.8 & U \\ 
13:52:33.60 & 30:59:38.2 &  39.6 &  39.6 & U \\ 
13:52:38.04 & 32:31:49.5 &  38.7 &  38.7 & U \\ 
13:52:49.41 & 32:13:00.8 &  43.9 &  61.9 & 25.9 \\ 
13:52:53.40 & 30:39:08.5 &  25.8 &  25.8 & U \\ 
13:52:54.50 & 29:32:36.8 &  41.3 &  41.3 & U \\ 
13:52:55.32 & 31:38:05.1 & 133.7 & 172.8 & 19.0 \\ 
13:52:56.75 & 29:25:58.4 & 133.3 & 133.3 & U \\ 
13:52:59.74 & 33:04:40.3 &  94.9 & 987.4 & 167.5 \\ 
13:53:07.07 & 33:36:22.7 &  51.0 & 187.8 & 62.0 \\ 
13:53:11.60 & 32:05:40.9 & 986.2 & 1046.2 & 15.5 \\ 
13:53:12.52 & 32:54:03.9 & 106.6 & 134.1 & 16.2 \\ 
13:53:13.35 & 31:12:45.0 &  42.7 &  42.7 & U \\ 
13:53:17.18 & 30:40:32.6 &  73.9 &  73.9 & U \\ 
13:53:19.15 & 31:09:06.2 &  73.0 &  73.0 & U \\ 
13:53:25.00 & 29:34:13.5 &  67.1 &  67.1 & U \\ 
13:53:27.54 & 32:18:21.2 & 355.3 & 413.6 & 8.3 \\ 
13:53:31.12 & 32:45:57.9 & 153.5 & 306.7 & 59.5 \\ 
13:53:37.94 & 31:33:21.9 &  39.7 &  39.7 & U \\ 
13:53:38.99 & 30:40:24.7 &  53.0 & 189.3 & 109.0 \\ 
13:53:40.02 & 29:26:35.7 &  82.3 &  82.3 & U \\ 
13:53:41.66 & 33:13:36.6 &  49.5 & 175.2 & 87.5 \\ 
13:53:43.30 & 30:51:56.6 & 138.6 & 242.4 & 51.8 \\ 
13:53:43.74 & 31:47:06.7 &  37.7 &  37.7 & U \\ 
13:53:44.66 & 33:04:39.8 &  49.3 & 104.9 & 46.8 \\ 
13:53:45.38 & 31:51:51.6 & 3207.9 & 5079.4 & 62.6 \\ 
13:53:47.44 & 30:05:50.7 &  47.0 &  47.0 & U \\ 
13:53:49.17 & 32:09:10.1 &  53.2 &  53.2 & U \\ 
13:53:50.97 & 29:56:19.1 &  70.4 &  70.4 & U \\ 
13:53:53.86 & 30:31:09.8 &  43.6 &  43.6 & U \\ 
13:53:56.31 & 29:33:07.3 & 289.9 & 378.9 & 29.5 \\ 
13:54:00.30 & 33:36:38.7 &  81.8 &  81.8 & U \\ 
13:54:00.73 & 30:56:27.2 &  77.2 & 138.1 & 57.4 \\ 
13:54:00.92 & 32:57:27.7 &  77.6 & 168.3 & 53.4 \\ 
13:54:05.23 & 31:39:01.8 & 1319.6 & 1623.7 & 46.2 \\ 
13:54:07.44 & 33:16:54.1 & 113.0 & 113.0 & U \\ 
13:54:08.85 & 30:48:34.7 &  72.0 &  72.0 & U \\ 
13:54:11.72 & 30:42:20.1 &  25.1 &  25.1 & U \\ 
13:54:12.81 & 30:18:45.5 & 115.4 & 115.4 & U \\ 
13:54:14.12 & 30:38:25.1 &  64.0 &  64.0 & U \\ 

\end{tabular}
\end{table}

\begin{table}
\begin{tabular}{l l c r r}
\hline
RA (h:m:s) & Dec (d:m:s) & S$_{\rm {peak}}$ & S$_{\rm {int}}$ & Size \\
\hline

13:54:15.39 & 30:34:06.6 &  53.4 &  53.4 & U \\ 
13:54:16.46 & 30:42:05.8 &  35.5 &  35.5 & U \\ 
13:54:25.97 & 31:52:17.8 &  37.4 &  37.4 & U \\ 
13:54:26.35 & 30:05:35.3 &  79.0 & 100.7 & 19.1 \\ 
13:54:30.62 & 33:13:45.1 &  39.3 &  39.3 & U \\ 
13:54:32.01 & 32:03:30.6 & 847.1 & 1703.2 & 106.2 \\ 
13:54:33.34 & 30:56:04.0 & 225.9 & 267.9 & 47.2 \\ 
13:54:33.74 & 29:57:32.4 & 416.0 & 512.7 & 52.1 \\ 
13:54:37.29 & 31:44:21.1 &  65.2 & 120.9 & 44.1 \\ 
13:54:39.32 & 31:51:51.9 &  36.4 &  36.4 & U \\ 
13:54:48.26 & 29:21:05.0 & 104.4 & 157.3 & 20.8 \\ 
13:54:50.43 & 33:24:15.3 &  51.1 & 243.0 & 59.3 \\ 
13:54:51.32 & 32:47:24.5 & 175.8 & 229.7 & 56.9 \\ 
13:54:51.43 & 32:31:58.1 &  89.2 &  89.2 & U \\ 
13:54:59.24 & 30:36:07.1 & 714.5 & 764.2 & 13.1 \\ 
13:55:01.25 & 29:43:17.1 & 592.9 & 714.9 & 10.9 \\ 
13:55:01.36 & 33:06:27.1 &  54.9 &  54.9 & U \\ 
13:55:04.58 & 31:32:12.6 &  31.3 &  31.3 & U \\ 
13:55:06.00 & 30:34:59.2 & 165.5 & 243.1 & 47.0 \\ 
13:55:06.07 & 31:51:31.1 &  90.8 &  90.8 & U \\ 
13:55:11.79 & 32:57:42.8 & 134.3 & 134.3 & U \\ 
13:55:11.91 & 32:56:12.7 & 109.1 & 109.1 & U \\ 
13:55:12.94 & 32:17:35.9 & 379.1 & 750.8 & 54.1 \\ 
13:55:13.15 & 32:51:59.0 &  33.1 &  33.1 & U \\ 
13:55:14.91 & 31:56:26.7 & 204.1 & 204.1 & U \\ 
13:55:16.45 & 29:21:54.3 & 102.7 & 366.0 & 91.2 \\ 
13:55:19.29 & 29:25:34.7 &  90.2 & 187.7 & 36.9 \\ 
13:55:21.78 & 31:44:39.2 & 240.3 & 869.9 & 64.8 \\ 
13:55:21.79 & 31:28:15.9 &  88.8 &  88.8 & U \\ 
13:55:22.69 & 30:34:32.2 &  89.8 &  89.8 & U \\ 
13:55:25.45 & 29:33:11.6 & 700.6 & 817.4 & 77.4 \\ 
13:55:26.94 & 32:34:58.6 &  92.0 &  92.0 & U \\ 
13:55:28.65 & 32:19:52.2 &  47.9 &  47.9 & U \\ 
13:55:30.13 & 30:31:30.2 & 175.9 & 195.2 & 13.9 \\ 
13:55:30.30 & 31:12:36.5 & 108.4 & 108.4 & U \\ 
13:55:31.93 & 32:48:28.1 &  44.8 &  44.8 & U \\ 
13:55:35.14 & 30:33:21.7 &  53.8 &  79.4 & 25.1 \\ 
13:55:35.89 & 29:54:00.8 & 131.7 & 131.7 & U \\ 
13:55:37.65 & 32:29:15.3 &  60.9 &  60.9 & U \\ 
13:55:39.20 & 29:32:19.6 &  77.7 &  77.7 & U \\ 
13:55:41.11 & 30:24:15.1 & 174.0 & 174.0 & U \\ 
13:55:46.12 & 32:38:04.6 & 566.6 & 649.8 & 20.2 \\ 
13:55:51.35 & 31:37:34.9 &  33.3 &  33.3 & U \\ 
13:55:56.72 & 33:29:06.5 &  46.8 &  46.8 & U \\ 
13:55:58.28 & 31:08:32.9 &  52.2 &  52.2 & U \\ 
13:55:58.79 & 31:19:39.7 & 172.3 & 192.9 & 15.1 \\ 
13:56:03.56 & 30:59:00.6 &  34.1 &  34.1 & U \\ 
13:56:04.82 & 31:45:07.6 & 107.7 & 107.7 & U \\ 
13:56:07.51 & 30:04:55.0 & 105.1 & 172.2 & 21.2 \\ 
13:56:08.03 & 31:53:05.2 & 792.1 & 846.2 & 9.9 \\ 
13:56:15.53 & 30:52:20.2 & 1037.0 & 1459.4 & 60.0 \\ 
13:56:15.67 & 30:06:40.4 & 114.0 & 114.0 & U \\ 
13:56:17.08 & 32:21:14.6 & 211.0 & 211.0 & U \\ 
13:56:19.79 & 32:03:53.5 &  58.8 &  58.8 & U \\ 
13:56:19.96 & 31:26:28.0 &  97.9 & 436.7 & 85.6 \\ 
13:56:24.73 & 31:17:06.4 & 173.5 & 282.7 & 39.5 \\ 
13:56:29.14 & 30:11:24.3 &  42.7 &  42.7 & U \\ 
13:56:30.33 & 29:41:19.8 & 811.5 & 1007.1 & 71.3 \\ 
13:56:36.64 & 30:36:17.5 &  29.5 &  29.5 & U \\ 
13:56:38.08 & 31:02:30.2 & 212.3 & 212.3 & U \\ 
13:56:39.46 & 30:16:51.3 &  57.2 &  57.2 & U \\ 
13:56:44.86 & 32:19:46.2 & 2869.4 & 8854.4 & 161.2 \\ 
13:56:46.56 & 32:53:48.0 &  51.0 &  51.0 & U \\ 
13:56:48.06 & 30:22:40.1 & 338.8 & 619.2 & 95.6 \\ 

\end{tabular}
\end{table}

\begin{table}
\begin{tabular}{l l c r r}
\hline
RA (h:m:s) & Dec (d:m:s) & S$_{\rm {peak}}$ & S$_{\rm {int}}$ & Size \\
\hline

13:56:50.60 & 32:04:47.9 & 176.5 & 291.9 & 48.7 \\ 
13:56:54.91 & 30:56:41.1 &  54.3 & 211.7 & 86.3 \\ 
13:57:00.91 & 30:24:46.7 & 132.6 & 201.7 & 31.0 \\ 
13:57:00.99 & 33:19:12.8 &  50.5 &  50.5 & U \\ 
13:57:02.35 & 31:54:57.9 &  48.8 &  48.8 & U \\ 
13:57:02.81 & 31:48:13.3 & 271.4 & 311.2 & 32.6 \\ 
13:57:02.95 & 31:11:51.9 &  27.0 &  27.0 & U \\ 
13:57:04.05 & 29:32:29.9 &  93.8 &  93.8 & U \\ 
13:57:08.64 & 32:58:21.7 & 120.5 & 283.1 & 47.9 \\ 
13:57:09.71 & 32:50:44.3 & 673.5 & 815.9 & 10.7 \\ 
13:57:13.88 & 31:03:46.1 & 232.7 & 362.1 & 20.6 \\ 
13:57:14.42 & 29:57:56.2 &  48.9 &  48.9 & U \\ 
13:57:20.42 & 33:11:31.2 &  52.6 &  52.6 & U \\ 
13:57:20.54 & 32:03:15.7 &  86.0 &  86.0 & U \\ 
13:57:22.13 & 30:41:55.5 & 1214.5 & 1470.0 & 32.6 \\ 
13:57:23.95 & 33:01:11.7 &  94.8 &  94.8 & U \\ 
13:57:26.29 & 29:58:27.9 & 103.7 & 162.0 & 26.7 \\ 
13:57:37.43 & 30:46:07.6 &  37.8 &  37.8 & U \\ 
13:57:41.43 & 30:09:06.4 &  51.8 &  51.8 & U \\ 
13:57:55.22 & 31:39:06.7 & 628.1 & 877.0 & 52.5 \\ 
13:58:01.75 & 32:37:31.3 &  55.2 &  55.2 & U \\ 
13:58:16.15 & 29:55:33.7 & 152.2 & 152.2 & U \\ 
13:58:19.54 & 31:18:35.4 & 1596.0 & 2819.0 & 141.1 \\ 
13:58:26.09 & 30:28:12.0 & 136.6 & 220.7 & 20.2 \\ 
13:58:26.41 & 29:55:44.2 &  58.0 &  58.0 & U \\ 
13:58:27.93 & 29:57:16.4 &  48.3 &  48.3 & U \\ 
13:58:27.94 & 30:31:48.4 &  75.1 &  75.1 & U \\ 
13:58:31.39 & 29:54:59.9 &  37.7 &  37.7 & U \\ 
13:58:31.88 & 32:22:35.4 &  41.7 &  41.7 & U \\ 
13:58:33.04 & 30:32:32.7 &  32.3 &  32.3 & U \\ 
13:58:42.27 & 31:00:47.6 &  47.6 &  47.6 & U \\ 
13:58:48.19 & 31:25:55.4 &  95.7 & 186.2 & 34.2 \\ 
13:58:49.23 & 32:21:51.9 &  46.2 &  46.2 & U \\ 
13:58:50.25 & 30:36:19.9 & 144.4 & 211.9 & 23.1 \\ 
13:58:53.28 & 30:26:22.7 & 246.4 & 433.4 & 34.9 \\ 
13:58:53.86 & 32:21:07.3 &  30.7 &  30.7 & U \\ 
13:59:00.94 & 31:30:59.7 &  78.0 & 115.6 & 23.2 \\ 
13:59:01.21 & 31:33:59.9 & 112.1 & 112.1 & U \\ 
13:59:06.16 & 29:54:29.8 &  85.9 & 125.2 & 23.9 \\ 
13:59:14.46 & 31:41:50.2 &  35.1 &  35.1 & U \\ 
13:59:18.09 & 30:25:02.5 & 114.5 & 270.4 & 21.1 \\ 
13:59:24.76 & 32:30:42.9 &  78.6 & 263.7 & 49.7 \\ 
13:59:26.41 & 31:51:20.4 &  40.1 &  40.1 & U \\ 
13:59:27.82 & 30:14:31.9 &  46.6 &  46.6 & U \\ 
13:59:32.18 & 30:22:01.0 &  41.1 &  41.1 & U \\ 
13:59:36.41 & 31:14:09.4 &  36.5 &  36.5 & U \\ 
13:59:37.44 & 30:06:44.5 &  38.0 & 115.9 & 61.0 \\ 
13:59:47.69 & 30:32:26.9 &  39.7 &  39.7 & U \\ 
13:59:53.27 & 32:25:31.4 & 110.0 & 110.0 & U \\ 
13:59:54.23 & 31:45:22.8 & 460.1 & 559.8 & 14.7 \\ 
14:00:06.35 & 30:36:17.2 &  47.3 &  47.3 & U \\ 
14:00:06.91 & 32:29:10.8 &  50.8 &  50.8 & U \\ 
14:00:10.86 & 31:23:29.0 &  68.5 & 112.7 & 29.4 \\ 
14:00:13.88 & 32:39:46.4 &  58.8 &  58.8 & U \\ 
14:00:18.08 & 32:42:32.7 &  43.0 &  71.3 & 35.5 \\ 
14:00:34.47 & 30:12:09.6 &  64.3 &  64.3 & U \\ 
14:01:05.43 & 30:20:18.2 &  73.2 &  73.2 & U \\ 
14:01:21.44 & 31:37:48.8 &  30.7 &  30.7 & U \\ 
14:01:22.31 & 31:05:08.1 & 132.7 & 243.1 & 26.5 \\ 
14:01:32.99 & 31:08:55.6 &  37.7 & 132.1 & 44.1 \\ 

\end{tabular}
\end{table}

\end{document}